\documentclass[unnumsec,webpdf,contemporary,large,svgnames]{oup-authoring-template}
\graphicspath{{Fig/}}

\pdfoutput=1 

\usepackage{natbib}
\setcitestyle{authoryear,round,comma}
\usepackage{xcolor}

\begin{document}

\copyrightyear{}
\appnotes{Applications Note}

\firstpage{1}

\title[iSEEtree]{iSEEtree: interactive explorer for hierarchical data}

\author[1,$\ast$]{Giulio Benedetti\ORCID{0000-0002-8732-7692}}
\author[1]{Ely Seraidarian\ORCID{0009-0008-8602-093X}}
\author[1]{Theotime Pralas\ORCID{0009-0004-2489-9601}}
\author[1]{Akewak Jeba\ORCID{0009-0007-1347-7552}}
\author[1]{Tuomas Borman\ORCID{0000-0002-8563-8884}}
\author[1]{Leo Lahti\ORCID{0000-0001-5537-637X}}

\authormark{Benedetti et al.}

\address[1]{\orgdiv{Department of Computing}, \orgname{Faculty of Technology, University of Turku}, \orgaddress{\street{Yliopistonmäki}, \postcode{FI-20014}, \state{Turku}, \country{Finland}}}

\corresp[$\ast$]{Corresponding author. \href{email}{giulio.benedetti@utu.fi}}

\abstract{
\textbf{Motivation:} Hierarchical data structures are prevalent across several fields of research, as they represent an organised and efficient approach to study complex interconnected systems. Their significance is particularly evident in microbiome analysis, where microbial communities are classified at various taxonomic levels along the phylogenetic tree. In light of this trend, the R/Bioconductor community has established a reproducible analytical framework for hierarchical data, which relies on the highly generic and optimised TreeSummarizedExperiment data container. However, using this framework requires basic proficiency in programming.\\
\textbf{Results:} To reduce the entry requirements, we developed iSEEtree, an R shiny app which provides a visual interface for the analysis and exploration of TreeSummarizedExperiment objects, thereby expanding the interactive graphics capabilities of related work to hierarchical structures. This way, users can interactively explore several aspects of their data without the need for extensive knowledge of R programming. We describe how iSEEtree enables the exploration of hierarchical multi-table data and demonstrate its functionality with applications to microbiome analysis.\\
\textbf{Availability and Implementation:} iSEEtree was implemented in the R programming language and is available on Bioconductor at \href{https://bioconductor.org/packages/iSEEtree}{https://bioconductor.org/packages/iSEEtree} under an Artistic 2.0 license.\\
\textbf{Contact:} \href{email}{giulio.benedetti@utu.fi} or \href{email}{leo.lahti@utu.fi}.}

\maketitle

\section{Introduction}

Interactive exploration plays a key role in data science. It can support early analytical stages, such as quality control and hypothesis generation, as well as more advanced ones, such as the final interpretation and reporting of the study results. However, the hierarchical nature of data in microbiome research and other fields necessitates specific analytical and visual solutions \citep{gloor2017microbiome}. Therefore, the R/Bioconductor community has developed a robust bioinformatic framework to effectively deal with hierarchical data \citep{core2024r, gentleman2004bioconductor, borman2024mia}. Despite these advances, there remains a considerable gap in the availability of tailored interactive tools to perform exploratory analysis of hierarchical data in fields such as microbiome research.

The current work has been motivated especially by emerging needs in modern microbiome analysis, where data structures are inherently hierarchical. Microbiome data is typically organised in abundance tables, whose rows and columns represent taxonomic features and experimental samples, respectively. Features vary in the degree of relatedness to other features, which leads to a hierarchical structure that is routinely described by phylogenetic trees. Given the complexity of hierarchical systems such as microbial communities, it is critical to carry a diverse and sound toolkit of analytical and visual solutions. To this end, iSEEtree was developed as an R shiny app that combines existing methods for hierarchical data analysis with the interactive graphic capabilities of the iSEE package \citep{chang2024shiny, rue2018isee}. This integration enables users to leverage code-based analytical and plotting functionality without the necessity for an extensive programming knowledge thereof.

Previously developed solutions for microbiome data exploration are either based on the phyloseq class (Shiny-phyloseq by \citet{mcmurdie2015shiny}) or on package-specific object types (animalcules by \citet{zhao2021animalcules}). While these are also valid approaches, they may lack either scalability to multi-table datasets or compatibility with other Bioconductor packages. In contrast, iSEEtree adopts the general-purpose TreeSummarizedExperiment (TreeSE) container, and thus inherits its multi-assay functionality and generalisability to a vast array of datasets with a hierarchical structure \citep{huang2020treesummarizedexperiment}. Among other advantages, this allows to explore multiple panels in parallel and dynamically transmit information between them.

Overall, by combining the strengths of existing software for hierarchical data analysis, iSEEtree offers flexible and user-friendly solutions for interactive data visualisation and exploration without the need for an extensive knowledge of programming, assisting new and more experienced practitioners in the analysis and discovery of hierarchical multi-table data.

\section{Software Implementation}

iSEEtree bridges the gap between effective visualisation of hierarchical data and user-friendly interactive analysis. However, several previous packages critically contributed to its development. In particular, this framework relies on the TreeSE container as its object type and derives its graphical and interactive capabilities from the mia and iSEE packages available in Bioconductor. In the following section, we discuss how iSEEtree complements and extends the existing ecosystem.

\begin{figure*}
  \includegraphics[width=\linewidth]{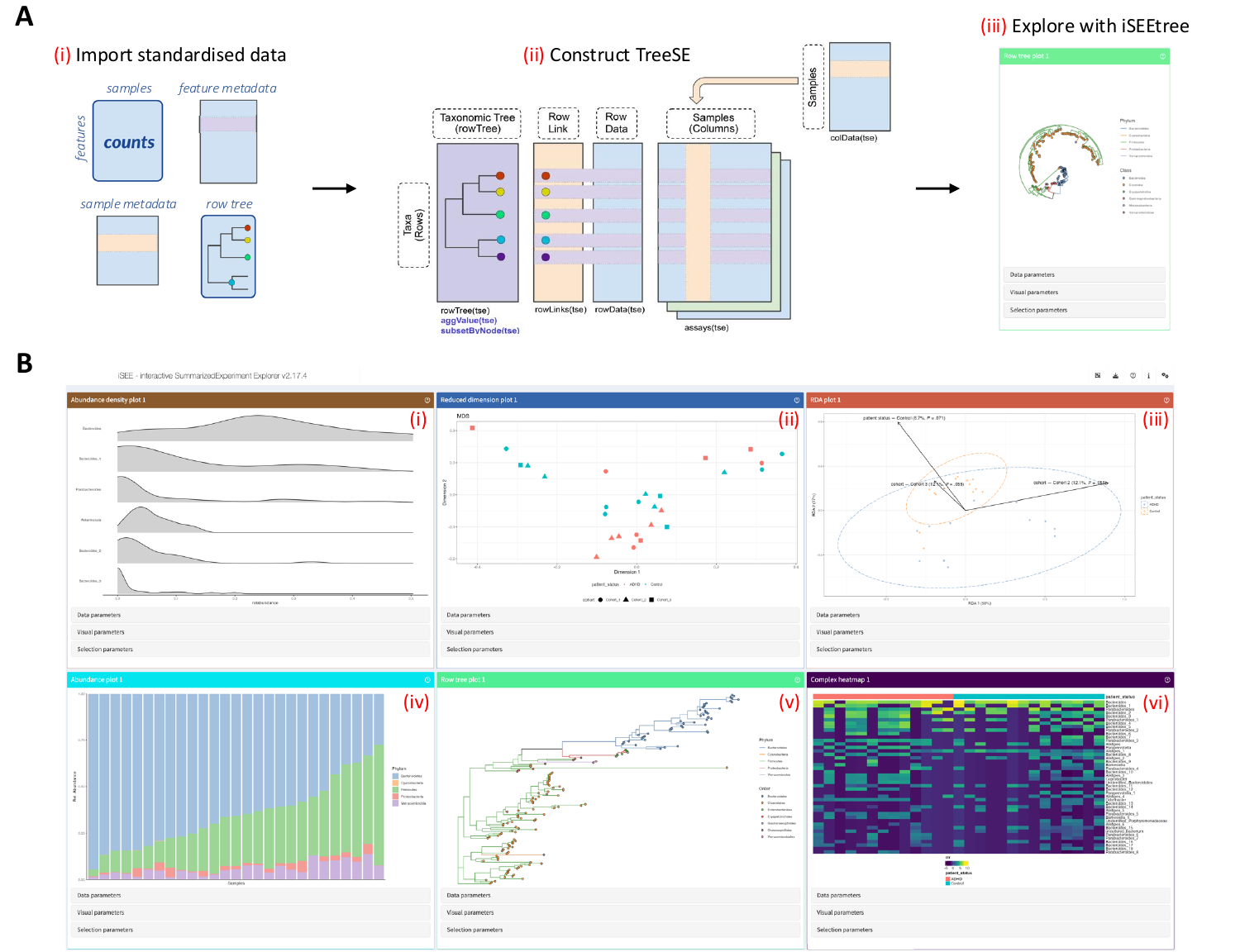}
  \caption{iSEEtree operative framework. \textbf{A}, Analytical workflow. (i) Data in standardised format is imported in R. This may include abundance assays and optional metadata on features, samples and tree hierarchies. (ii) A TreeSE object is constructed from the data. This is an optimised and standardised container for hierarchical data analysis. (iii) The object is input to iSEEtree, which launches an interface with a set of data-dependent visualisations. Adapted from \citet{huang2020treesummarizedexperiment}. \textbf{B}, iSEEtree panel layout. The app provides a customisable interface, where each panel describes a certain aspect of the data under investigation. Panels can help explore the compositional organisation (i, iv and vi), identify main components (ii and iii) and inspect the hierarchical structure (v) of the data. The dataset shown in the snapshot was derived from \citet{tengeler2020gut}.}
  \label{fig:framework}
\end{figure*}

\subsection{Data containers}

Data containers constitute a standardised and optimised framework to store and manipulate data, ensuring that information is systematically organised and readily accessible. In this context, the TreeSE container is designed to expand the functionality of its superclasses, the SummarizedExperiment and SingleCellExperiment, to hierarchical structures \citep{huang2020treesummarizedexperiment, martin2024se, amezquita2020orchestrating}. TreeSE objects can host multiple numeric tables as assays along with metadata on samples (columns) and features (rows) as colData and rowData, respectively. Additionally, results from dimensionality reduction methods, such as Principal Component Analysis (PCA) and Redundancy Analysis (RDA), can be stored in the corresponding reducedDim slot, which was introduced by and is inherited from the SingleCellExperiment class. However, what ultimately distinguishes TreeSE from other containers is its ability to account for the hierarchical structure of the data through the rowTree and colTree elements, which can accommodate phylogenetic trees or sample hierarchies and map to features or samples of the assays by means of rowLinks or colLinks, respectively.

The flexibility and self-contained nature of the TreeSE container has led several microbiome analysis frameworks in Bioconductor to adopt it as their base class. For instance, the mia family of packages provide a range of tools to perform transformations, statistical analysis and visualisation on TreeSE data objects. Building on this foundation, iSEEtree leverages the advantages of such container to facilitate the interactive exploration of hierarchical data.

\subsection{Operative framework}

The workflow to use iSEEtree can be divided into three steps (Fig. \ref{fig:framework}A). First, data is imported from standardised file formats and processed according to one's visualisation needs. This may include multiple abundance assays supplemented with metadata on features, samples and tree hierarchies. Alternatively, ready-to-use datasets are also available to experiment with the app functionality.  Next, a TreeSE object is constructed from the data. This step often takes place automatically when using importers for standardised file formats. Common methods to import and manipulate TreeSE objects are described in the online book "Orchestrating Microbiome Analysis" \citep{lahti2021orchestrating}. Finally, the app can be launched from an R console or RStudio by executing the command \texttt{iSEE(tse)}, where the input object is of type TreeSE. If the input data belongs to a more generic class, such as SummarizedExperiment, the panel layout will instead follow the method defined in the parent package iSEE.

After launching the app, an interface with a predefined yet customisable set of panels is created, where each panel showcases a certain aspect of the data under investigation (Fig. \ref{fig:framework}B). For each panel, several parameters are available to transform data (\textit{Data parameters}), control aesthetics (\textit{Visual parameters}) and select a subset of the data (\textit{Selection parameters}). Their usage is described in the tours accessible by clicking on the question marks. Once the desired visualisation is achieved, plots can be downloaded to report and publish findings.

\subsection{App functionality}

The current version of iSEEtree provides the tools to perform three different stages of hierarchical data analysis. First, preliminary compositional exploration gives an overarching perspective on the structure and organisation of the system. For this purpose, the Abundance and Abundance density plots display sample-wise composition by means of a barplot and a density plot, respectively. In addition, the Complex heatmap plot inherited from iSEE can be used to visualise assay counts for multiple features at a time. Second, supervised and unsupervised ordination techniques estimate diversity across samples and determine the main biological or technical factors that contribute to it. Such information can be explored with the Reduced dimension and RDA plots, which showcase the results of dimensionality reduction from the reducedDim slot, and the Loading plots, which presents features ordered by the importance of their contribution to a reduced dimension of choice. Last but not least, the investigation of hierarchical structures reveals how features or samples relate to one another and can be used to identify dominant groups within the overall system. Such aspects can be visualised with the Row tree and Column tree plots, which represent the relational organisation of the features or the samples of a system by means of a tree. This support for hierarchical data is a unique feature of iSEEtree that distinguishes it from other extensions of the iSEE package.

iSEEtree is built on top of the independently developed package iSEE, which provides panels for the generic SummarizedExperiment class, a container to organise data in multiple tables supplemented with metadata on samples and features \citep{rue2018isee, huang2020treesummarizedexperiment}. Consequently, our app inherits the complete array of primary and auxiliary functionality from this foundation, including multiple interactive plots and tables, selectable parameters, dynamic linking between panels, interactive panel tours, custom panel colouring, code tracking for reproducibility and downloadable results. Panels derived from iSEE are not only supported in our work, but often play a critical role in the exploration of hierarchical data. Examples include the Row data and Column data tables, designed to easily list and select sets of features and samples, or the Complex heatmap and Reduced dimension plots, meant to visualise assays at full or reduced dimensionality. However, certain panels require the presence of specific elements in the TreeSE object, and they will appear only when these are available.

\section{Applications}

\subsection{Use cases}

The basic usage of iSEEtree is demonstrated for \citet{tengeler2020gut}, a study on the effects of gut microbiome on attention-deficit/hyperactivity disorder (ADHD) in humanised mice. The dataset contains 151 features from 27 stool samples obtained with 16rRNA gene sequencing. The tutorial guides users through importing and preparing the dataset for visualisation, launching the app and customising the initial panel layout. Fig. \ref{fig:framework}B depicts a snapshot of the app for this dataset after adjusting the panel parameters. This tutorial can be found in the online package documentation.

\subsection{Extensions}

Similar visualisation approaches for hierarchical data are found across many research fields. However, the preprocessing and analytical methods for such data largely depend on the objective of the study, which often varies between different fields. In this respect, iSEEtree provides highly generic panels for hierarchical data of any origin, while domain-specific applications may be subsequently developed on its foundation. A notable example of this integration is illustrated by miaDash, a web app for the interactive analysis and exploration of microbiome data \citep{benedetti2024dash}. This online extension enables users to import, manipulate and ultimately visualise hierarchical datasets according to the unique requirements of microbiome analysis, all through a fully graphical interface built on the base of iSEEtree. To summarise, iSEEtree not only serves as a flexible tool to explore hierarchical data across several domains, but also paves the way for tailored implementations such as miaDash. More information about this web app can be found in the online package documentation.

\section{Limitations}
The three major limitations of iSEEtree pertain to scalability, method coverage and accessibility. First, the app may significantly slow down as data size increases. While we are exploring ways to optimise performance, users can speed up calculations by reducing the size of the data (e.g. through subsetting or agglomerating). Second, the current panel catalogue is relatively restricted and parameters do not cover all options available in the corresponding command line environment. Although the R scripts generated by the app can be modified and extended, we also aim to expand the selection of iSEEtree panels and add more interactive features to the existing ones. Third, it is still necessary to build TreeSE objects and launch the app either from R or RStudio. Because this may represent a barrier for novice users, an entirely graphical interface could be created by incorporating the generic panels of iSEEtree with custom dashboards for domain-specific data analysis. One example of such integration is illustrated by the miaDash app in the context of microbiome research. In summary, enhanced scalability, method coverage and accessibility can further improve the quality of the app and enrich the overall user experience.

\section{Conclusion}

Although the study of hierarchical data has become a routine practice in statistical analysis, a certain degree of programming skills is generally required to perform even the most basic evaluations. To address this need, we developed iSEEtree, a publicly available R shiny app that provides a flexible and user-friendly graphical interface to explore hierarchical data. The app was demonstrated on a typical microbiome dataset. However, its functionality can be extended to any domain which deals with hierarchical data. iSEEtree is part of the R/Bioconductor ecosystem, supporting contributions from the open developer community. By providing a graphical interface with generic visual solutions, the app assists novice and experienced practitioners from various fields, regardless of their knowledge of programming, to effectively navigate hierarchical multi-table data and vividly report findings from their studies, ultimately promoting reproducible practices for scientific discovery in microbiome research and other application domains.

\section{Software Availability}
The iSEEtree package is available on Bioconductor at \href{https://bioconductor.org/packages/iSEEtree}{https://bioconductor.org/packages/iSEEtree} under an Artistic 2.0 license. Package functionality and example use cases are further described in the online documentation at \href{https://microbiome.github.io/iSEEtree/}{https://microbiome.github.io/iSEEtree/}. This article reported iSEEtree v1.0, which requires R v4.4.0 or above and other dependencies from Bioconductor v3.20 or above.

\section{Competing interests}
No competing interest is declared.

\section{Author contributions statement}
All authors read and approved the final version of the manuscript. G.B., T.B and L.L conceptualised the software; G.B. and E.S. created the package; T.B, E.S, T.P and G.B. developed underlying visualisation methods. A.J. and G.B. containerised the package into a web app. G.B., T.B. and L.L. authored and reviewed the manuscript.

\section{Acknowledgments}
This project has received funding from the European Union’s Horizon 2020 research and innovation programme (grant 952914) and from the Research Council of Finland (grant 330887). Special thanks are reserved to Kevin-Rue Albrecht and the other developers of the iSEE package for regular support and key technical advice.

\bibliography{bibliography}

\end{document}